\documentclass[epsfig,aps,prl,twocolumn,superscriptaddress,amsmath,amssymb]{revtex4}

\usepackage{dcolumn}
\usepackage{bm}
\usepackage[english]{babel}

\usepackage{ifpdf} 
\ifpdf

\usepackage[pdftex]{graphicx,color} 
\usepackage{epstopdf}  
\else

\fi

\usepackage{epsfig,graphicx}
\usepackage{float}

\begin{document}
\title{Nanophotonic control of the F{\"o}rster resonance energy transfer efficiency }

\author{Christian Blum}
\affiliation{Nanobiophysics (NBP), MESA+ Institute for Nanotechnology, University of Twente, 7500 AE Enschede, The Netherlands}
\altaffiliation{Email: c.blum@utwente.nl, URL: www.utwente.nl/tnw/ nbp, www.photonicbandgaps.com}

\author{Niels Zijlstra}
\affiliation{Nanobiophysics (NBP), MESA+ Institute for Nanotechnology, University of Twente, 7500 AE Enschede, The Netherlands}

\author{Ad Lagendijk}
\affiliation{Complex Photonic Systems (COPS), MESA+ Institute for Nanotechnology, University of Twente, 7500 AE Enschede, The Netherlands}
\affiliation{FOM-Institute AMOLF, Science Park, 1098 XG Amsterdam, The Netherlands}

\author{Martijn Wubs}
\affiliation{Department of Photonics Engineering, Technical University of Denmark, DK-2800 Kgs.~Lyngby, Denmark}

\author{Allard P. Mosk}
\affiliation{Complex Photonic Systems (COPS), MESA+ Institute for Nanotechnology, University of Twente, 7500 AE Enschede, The Netherlands}

\author{Vinod Subramaniam}
\affiliation{Nanobiophysics (NBP), MIRA Institute for Biomedical Engineering and Technical Medicine and MESA+ Institute for Nanotechnology, University of Twente, 7500 AE Enschede, The Netherlands}

\author{Willem L. Vos}
\affiliation{Complex Photonic Systems (COPS), MESA+ Institute for Nanotechnology, University of Twente, 7500 AE Enschede, The Netherlands}
\altaffiliation{URL: www.photonicbandgaps.com}

\date{Prepared July 5th, 2012}

\begin{abstract}
We have studied the influence of the local density of optical states (LDOS) on the rate and efficiency of F{\"o}rster resonance energy transfer (FRET) from a donor to an acceptor. 
The donors and acceptors are dye molecules that are  separated by a short strand of double-stranded DNA. 
The LDOS is controlled by carefully positioning the FRET pairs near a mirror.
We find that the energy transfer efficiency changes with LDOS, and that, in agreement with theory, the energy transfer rate is independent of the LDOS, which allows one to quantitatively control FRET systems in a new way.  
Our results imply a change in the characteristic F{\"o}rster distance, in contrast to common lore that this distance is fixed for a given FRET pair. 

\end{abstract}
\maketitle

In the field of cavity quantum electrodynamics it is recognized that the fundamental interactions between a single two-level quantum emitter - such as an atom, molecule, or quantum dot - and the light field can be exquisitely controlled by the photonic nano-environment of the emitter~\cite{Drexhage1970, Yablonovitch1987, Haroche92, Novotny06, Leistikow11}. 
In comparison, reports on the control of multiple interacting emitters, even in case of only two emitters, are scarce.
A particularly interesting and well-known optical emitter-emitter interaction is F{\"o}rster resonance energy transfer (FRET), which is a near-field nonradiative energy transfer between pairs of dipoles where the quantum of excitation energy is transferred from one emitter, called donor, to a second emitter, called acceptor~\cite{Forster1948}. 
FRET is the dominant energy transfer mechanism between emitters in nanometer proximity and plays a pivotal role in the photosynthetic apparatus of plants and bacteria~\cite{Grondelle1994, Scholes2003}. 
Many applications are based on FRET, ranging from photovoltaics~\cite{Chanyawadee2009,Buhbut2010} and  lighting~\cite{Baldo2000,Vohra2010}, to sensing~\cite{Medintz2003} where molecular distances~\cite{Stryer1978,Schuler2002} and interactions are probed~\cite{Garcia2004, Carriba2008}. 
The dipole-dipole interactions in FRET are also relevant to the storage and transfer of quantum information~\cite{John1991, Barenco1995, Lovett2003, Reina2004, Unold2005}.

The properties of a FRET system are traditionally controlled by the spectral properties of the coupled emitters, by their
distance $R$ with a typical $(R_{0}/R)^6$ dependence (where $R_{0}$ is the F{\"o}rster distance), or by the relative orientations of the dipole moments~\cite{Forster1948, Lakowicz2006}. 
It is an open question, however, whether F{\"o}rster transfer can be controlled purely by means of the photonic environment while leaving the FRET pair geometrically and chemically unchanged. 
The photonic environment is characterized by the local density of optical states (LDOS) that counts the number of photon modes available for emission, and is interpreted as the density of vacuum fluctuations~\cite{Sprik1996,Barnes1998}. 
While qualitative photonic effects on FRET have been reported~\cite{Kolaric2007,Yang2008}, there is an unresolved debate on how the F{\"o}rster energy transfer rate depends on the LDOS. 
Pioneering work by Andrew and Barnes suggested that the transfer rate depends linearly on the donor decay rate and thus the LDOS at the donor emission frequency~\cite{Andrew2000}, as was confirmed elsewhere~\cite{Nakamura2005}. 
While this result was supported by theory~\cite{Dung2002}, it was also stated that the transfer rate is differently affected by the LDOS. 
Subsequent work suggested a dependence on the LDOS squared~\cite{Nakamura2006}, or a transfer rate \emph{independent} of the LDOS~\cite{Dood2005}. 
Possible reasons for the disparity in these observations include lack of control on the donor-acceptor distance and on exact pairing of every donor to one acceptor, or cross-talk between neighboring FRET pairs. 
Therefore, we have decided to embark on a study of the relation between F{\"o}rster transfer and the LDOS, using precisely-defined, isolated, and efficient donor-acceptor pairs.
A precise control over the LDOS is realized by positioning the FRET pairs at well-defined distances to a metallic mirror~\cite{Drexhage1970,Barnes1998,Chance1978}. 

We obtain the energy transfer rate $\gamma_{\rm FRET}$ and concomitant efficiency $\eta_{\rm FRET}$ from measurements of the donor emission rate $\gamma_{\rm DA}$ in presence of the acceptor and the rate $\gamma_{\rm D}$ in absence of the acceptor~\cite{Lakowicz2006}. 
Adding a FRET acceptor to a donor introduces the energy transfer as an additional decay channel, hence the total decay rate $\gamma_{\rm DA}$ of a FRET-coupled donor equals $\gamma_{\rm DA} = \gamma_{\rm D} + \gamma_{\rm FRET}$. 
Therefore we obtain the energy transfer rate $\gamma_{\rm FRET}$ from 
\begin{equation}\label{gammaFRET}
\gamma_{\rm FRET} = \gamma_{\rm DA} - \gamma_{\rm D}.
\end{equation}
Since the decay rate $\gamma_{\rm D}$ of an isolated donor is equal to the sum of the radiative rate $\gamma_{\rm rad}$ and the nonradiative rate $\gamma_{\rm nr}$ ($\gamma_{\rm D} = \gamma_{\rm rad} + \gamma_{\rm nr}$), the energy transfer efficiency $\eta_{\rm FRET}$ is equal to 
\begin{equation}\label{etaFRET}
\eta_{\rm FRET} = \dfrac{\gamma_{\rm FRET}}{\gamma_{\rm FRET}+\gamma_{\rm rad}+\gamma_{\rm nr}} = 1 - \frac{\gamma_{\rm D}}{\gamma_{\rm DA}}.
\end{equation}
%

%
\begin{figure}
\centering
\includegraphics[width=1.0\columnwidth]{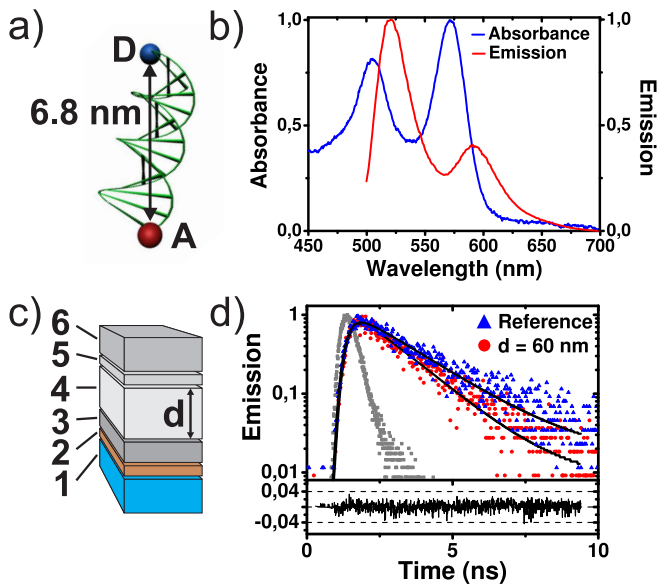}
\caption{(color)
a) Our FRET system consists of fluorescent dyes as donor (D, Atto488) and acceptor (A, Atto565) spaced by 6.8 nm by a short DNA strand.
b) Normalized absorbance (blue) and emission spectra (red) of the FRET system in solution.
c) Scheme of the sample to control the LDOS.
1: Si-wafer; 
2: adhesive layer; 
3: Ag mirror; 
4: SiO$_{2}$ spacer layer of precisely-defined thickness $d$ ($60 < d < 270$ nm); 
5: thin $< 20$ nm PVA film containing the FRET pairs; 
6: thick PMMA cover layer;
d) Typical normalized donor decay curves for the reference FRET sample ($d = \infty$, blue triangles) with $\gamma_{\rm DA}^0 = 0.555\pm 0.009$ ns$^{-1}$ and near a mirror $d = 60$ nm (red circles) with $\gamma_{\rm DA} = 0.704\pm 0.008$ ns$^{-1}$. 
Light grey squares: instrument response function, solid curves: single-exponential fits plus background. 
The residuals for the reference decay are random (bottom). }
\label{FRETscheme}
\end{figure}
%

The FRET pair we used consists of the efficient fluorescent dye molecules Atto488 as donor and Atto565 as acceptor emitters. 
The dye molecules are covalently bound via short linkers to the opposite ends of a 15 basepair long double-stranded DNA. 
The DNA forms a rigid helix that separates the donor and the acceptor by a precisely defined distance~\cite{Rindermann2011} 
of $6.8$ nm, see Fig.~\ref{FRETscheme}(a). 
The FRET system was synthesized with high purity to ensure that each donor was accompanied by an acceptor, since uncoupled FRET donor emitters would disturb the measurements. 
For intentional donor-only samples we used Atto488 covalently attached to an identical double strand of DNA (Atto488-DNA). 
The absorbance and emission spectra of the FRET system in solution show the relevant peaks of both molecules, see Fig.~\ref{FRETscheme}(b). 
The spectral overlap and separation between the two dyes ensures  F{\"o}rster transfer~\cite{Forster1948}.
Indeed, excitation of Atto488 yields emission from both Atto488 around $\lambda = 525$ nm and Atto565 around $\lambda = 590$ nm. 

To control the distance $d$ between the emitters and the silver mirror, and thus the LDOS the emitters experience, we fabricated SiO$_{2}$ spacer layers with thicknesses between 60 nm and 270 nm on the mirror, see Fig.~\ref{FRETscheme}(c). 
The emitters were deposited in a thin film ($< 20 $ nm) of PVA on top of the spacer layer, resulting in a random orientation of the donor-only or coupled donor-acceptor FRET systems with respect to the mirror. 
The emitter layer was covered with a thick layer of refractive-index matching PMMA to suppress interference effects from the sample-air interface. 
In addition, we fabricated reference samples without mirror, corresponding to a mirror at infinite distance ($d \rightarrow \infty$) ~\cite{Cesa2009}.

Decay curves were measured using a time-correlated single-photoncounting based lifetime imaging system~\cite{footnote:EPAPS}. 
The detection was limited to a spectral band around the donor emission peak ($\lambda =525 \pm 8$ nm) where only the donor emits.
Typical decay curves of the FRET donor are shown in Fig.~\ref{FRETscheme}(d). 
All curves could be well modeled using single exponential decays. 
The presence of unwanted donor fluorophores not coupled to a FRET acceptor fluorophore would result in a second decay component. 
Non-uniform distances and orientations between the FRET fluorophores would also lead to non single exponential decay. 
Therefore we conclude from the absence of additional decay components that our samples consist of the intended one donor on one acceptor FRET-pairs placed at a well-defined distance from the mirror. 

\begin{figure}
\centering
\includegraphics[width=1.0\columnwidth]{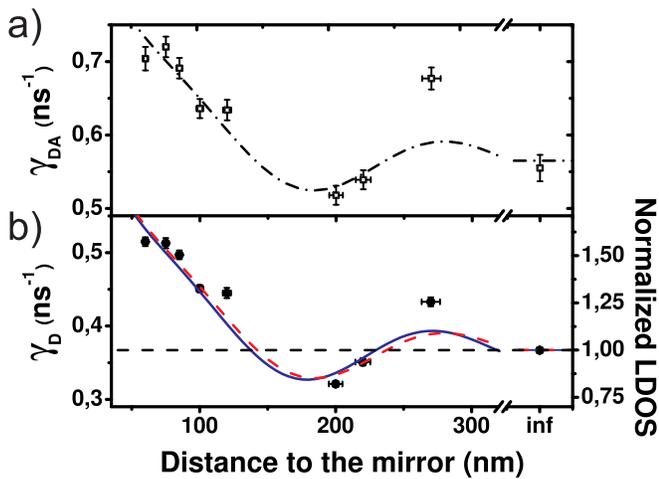}
\caption{(color)
a) Total donor decay rate $\gamma_{\rm DA}$ of the FRET samples oscillates with distance to the mirror.
Dash-dotted curve: model consisting of the calculated LDOS, averaged over the whole donor emission band, plus a constant FRET rate $\gamma_{\rm FRET}$  = 0.197 $\pm$ 0.006 ns$^{-1}$.
b) The total decay rate $\gamma_{\rm D}$ of the donor-only samples shows the well-known oscillation as a function of distance $d$ to the mirror. 
Solid blue curve: calculated LDOS for the center of the detection band at 525 nm; dashed red: calculated LDOS, averaged over the whole donor emission band.} 
\label{decaycurve_measured}
\end{figure}

We have recorded many decay curves from different areas of the sample, yielding an average of 1200 decay rates at each sampled emitter-mirror distance. 
From all decay rates we determined the decay rate distribution at each LDOS value, from which we extracted the most frequent decay rate and the width of the distribution.
In Fig.~\ref{decaycurve_measured} we present the resulting donor decay rates for the different emitter-mirror distances. 
For the donor-only samples we observe the well known oscillation of the decay rate $\gamma_{\rm D}$ that originates from the modification of the LDOS by the metallic mirror, see Fig.~\ref{decaycurve_measured}(b)~\cite{Drexhage1970}. 
The measured decay rates $\gamma_{\rm D}$ generally agree very well with the expected values based on the quantum efficiency determined for Atto488-DNA (see~\cite{footnote:EPAPS}), the peak emission wavelength of 525 nm for Atto488 and an isotropic distribution of emitter orientations. 
We consider the rate obtained for $d$ = 270 nm to be an outlier, probably originating from an unknown error in the spacer layer fabrication that may have resulted in deviations of the refractive index of the spacer layer or in the introduction of a contaminant that quenched the emitters.
The reference decay rate in the limit $d \rightarrow \infty$ was obtained from the samples without mirror~\cite{footnote:reference}. 
The good agreement between the modeled and the measured donor-only decay rates $\gamma_{\rm D}$ confirms that we achieve a precise control over the LDOS~\cite{Cesa2009}. 

\begin{figure}[!tbp]
\centering
\includegraphics[width=1.0\columnwidth]{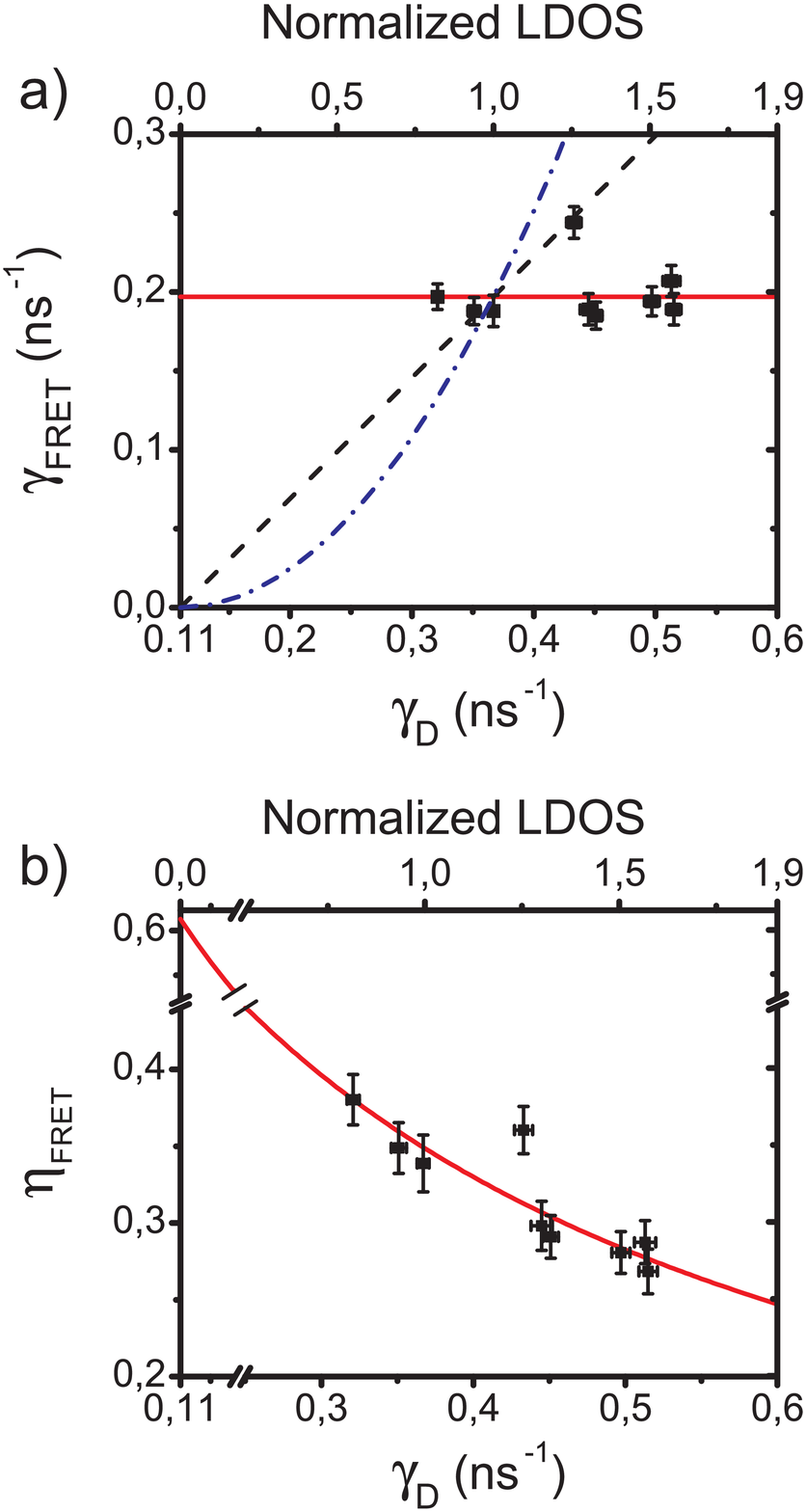}
\caption{(color)
a) F{\"o}rster energy transfer rate $\gamma_{\rm FRET}$ as a function of the donor-only decay rate $\gamma_{\rm D}$ that is proportional to the normalized LDOS at the donor emission frequency. 
The curves show the calculated FRET rates $\gamma_{\rm FRET}$ for proposed models: 
independent from LDOS (red, solid lines), linear (black, dash-dotted), and quadratic (blue, dashed) dependence on LDOS. 
The curve coincide for the reference case, at normalized LDOS = 1 ($d \rightarrow \infty$).
b) F{\"o}rster energy transfer efficiency $\eta_{\rm FRET}$ as a function of donor-only decay rate $\gamma_{\rm D}$ and LDOS.
The energy transfer efficiency agrees well with the model (red line), calculated from the rates $\gamma_{\rm rad}$, $\gamma_{\rm nr}$ and $\gamma_{\rm FRET}$ and the dependence of $\gamma_{\rm rad}$ on LDOS.
The maximum energy transfer efficiency occurs when the LDOS and $\gamma_{\rm rad}$ vanish, \emph{i.e.}, in a 3D photonic bandgap. 
Then $\gamma_{\rm D}$ equals $\gamma_{\rm nr}$ (here $0.109\pm 0.029$ ns$^{-1}$). }
\label{Forster_rate_eff}
\end{figure}

For the FRET samples we find a similar oscillation of decay rates $\gamma_{\rm DA}$ with distance to the mirror, as depicted in Fig.~\ref{decaycurve_measured}(a).
The decay rates $\gamma_{\rm DA}$ are always higher than the corresponding rates $\gamma_{\rm D}$ for the donor-only samples, which confirms that the additional decay rate $\gamma_{\rm FRET}$ due to energy transfer is positive (cf. Eq.~\ref{gammaFRET}), as is physically expected. 

To visualize the effect of the LDOS on F{\"o}rster transfer, we show in Fig.~\ref{Forster_rate_eff} as our main result the energy transfer rate $\gamma_{\rm FRET}$ and the concomitant efficiency $\eta_{\rm FRET}$ versus the donor-only decay rate $\gamma_{\rm D}$. 
According to Fermi's golden rule, $\gamma_{\rm D}$ depends linearly on the LDOS at the donor emission frequency~\cite{Novotny06}. 
We determined $\gamma_{\rm FRET}$ and $\eta_{\rm FRET}$ from the measured rates $\gamma_{\rm DA}$ and $\gamma_{\rm D}$ using Eqs.~(\ref{gammaFRET}) and~(\ref{etaFRET}) respectively. 
Apart from the point at normalized LDOS = 1.25 that originates from the anomalous $d = 270$ nm sample identified above, our data in Fig.~\ref{Forster_rate_eff}(a) show that the FRET rate $\gamma_{\rm FRET}$ is independent of the LDOS at the emission frequency of the donor, in agreement with De Dood \emph{et al}.~\cite{Dood2005}. 
The independence is confirmed by the observation that the average of the rates is $\gamma_{\rm FRET} = 0.197 \pm 0.006$ ns$^{-1}$, which agrees very well with the separate measurement $\gamma_{\rm FRET}^{0}= 0.188 \pm 0.010$ ns$^{-1}$ on the reference sample. 
We have also verified that $\gamma_{\rm FRET}$ is independent of the LDOS averaged over the emission band of the donor (cf. Fig.~\ref{decaycurve_measured}). 
Our data are not in agreement with a linear or quadratic dependence of the F{\"o}rster energy transfer rate on the LDOS  proposed earlier~\cite{Andrew2000,Nakamura2006}. 
Our observations are consistent with theoretical quantum-optical considerations, wherein the properties of the Green function are analyzed~\cite{Dung2002, footnote:EPAPS}. 
The Green function that describes the FRET mechanism can be expressed as an integral of the LDOS over a very broad frequency interval, from zero to about ten optical frequencies, plus an integral over all even higher frequencies~\cite{footnote:EPAPS}. 
In the former integral LDOS enhancements in certain frequency bands will be canceled by LDOS reductions in other frequency ranges, moreover the latter integral is the dominant contribution. 
Thus controlling the LDOS at optical frequencies does not significantly affect the F{\"o}rster energy transfer rate.

Turning from rates to efficiencies, we show in Fig.~\ref{Forster_rate_eff}(b) the FRET efficiency $\gamma_{\rm FRET}$ as a function of LDOS in the donor emission band. 
We observe clear changes of the FRET efficiency: for the range of LDOS sampled (0.86 to 1.65), the FRET efficiency is manipulated from $0.27$ to $0.38$, a relative change of over $30\%$ by purely nanophotonic means. 
The role of the LDOS on $\eta_{\rm FRET}$ can be perceived with the aid of Eq.~(\ref{etaFRET}): since the LDOS controls the radiative rate $\gamma_{\rm rad}$, it alters the competition between the different decay processes. 
The FRET efficiency $\eta_{\rm FRET}$ can be reduced to effectively zero by greatly enhancing $\gamma_{\rm rad}$ to the point that it dominates the FRET and nonradiative rates. 
On the other hand, the energy transfer efficiency can be increased by reducing the LDOS and thus $\gamma_{\rm rad}$. 
The maximum achievable transfer efficiency is reached in the limit of vanishing LDOS and $\gamma_{\rm rad}$, which corresponds experimentally to a photonic crystal with a complete 3D photonic band gap~\cite{Yablonovitch1987, Leistikow11}. 
The maximum achievable transfer efficiency is limited by the nonradiative rate to $\eta_{\rm FRET}^{\rm max} = \gamma_{\rm FRET}/(\gamma_{\rm nr}+\gamma_{\rm FRET})$. 
For the FRET system studied here, the maximum energy transfer efficiency is $64\%$, see Fig.~\ref{Forster_rate_eff}(b). 
In the intriguing situation of a FRET donor with unit quantum efficiency placed in a 3D photonic bandgap, the energy transfer efficiency will approach $100 \%$, independent of the acceptor properties, and the distances and orientations between the FRET partners. 
Thus a 3D photonic band gap is a promising platform to shield dipole-dipole interactions for quantum information applications.

In summary, our observation that the F{\"o}rster energy transfer rate between donors and acceptors with a fixed separation and in a well-defined geometry is independent of the LDOS settles a debate in the literature. 
As a consequence, the energy transfer efficiency $\eta_{\rm FRET}$ in a FRET system can be nanophotonically modified without changing the FRET system itself. 
The change in $\eta_{\rm FRET}$ implies a change in the characteristic F{\"o}rster distance $R_0$, which is defined as the distance at which $\eta_{\rm FRET}$ = $50\%$ for a given donor-acceptor pair~\cite{Lakowicz2006}.
Using the known distance between the FRET donor and acceptor together with the values determined for $\eta_{\rm FRET}$ in our experiment, we calculate $R_0$ to change from 6.1 nm for the reference sample to values between 5.7 nm and 6.3 nm by changing the LDOS. 
Our result that the characteristic F{\"o}rster distance $R_0$ can be changed via the LDOS is counterintuitive for the biochemical literature, where $R_0$ is taken to be fixed by the molecular properties of the FRET system~\cite{Lakowicz2006}.
In this sense, control over the LDOS allows one to change the scale of the molecular ruler when FRET is used to probe nanometer distances or molecular interactions.
Moreover, the basic understanding of nanophotonic control of the FRET efficiency can be also exploited in areas such as lighting technology, where FRET may form an unwanted loss channel, and in advanced photovoltaic technology, where FRET is an important mechanism to transport energy away from the capture site, in analogy to biological photosynthesis.

We thank Hannie van den Broek for sample preparation, Cock Harteveld for experimental help, Yanina Cesa and Danang Birowosuto for contributions early on, and Bill Barnes for discussions.
This work is part of the research programs of FOM and STW that are financially supported by NWO. APM acknowledges a Vidi grant from NWO and European Research Council grant 279248. NZ acknowledges support from NWO Chemical Sciences TOP grant number 700.58.302 to VS.
WLV thanks Smartmix Memphis.


\begin{thebibliography}{}

\bibitem{Drexhage1970}
K.H. Drexhage,
J. Lumin. \textbf{1-2}, 693 (1970).

\bibitem{Yablonovitch1987}
E. Yablonovitch,
Phys. Rev. Lett. \textbf{58}, 2059 (1987).

\bibitem{Haroche92}
S. Haroche, in \emph{Fundamental systems in quantum optics},
Eds. J. Dalibard, J.M. Raimond, and J. Zinn-Justin
(North Holland, Amsterdam, 1992), p. 767.

\bibitem{Novotny06}
L. Novotny and B. Hecht, 
\emph{Principles of Nano-Optics}
(Cambridge University Press, Cambridge, 2006).

\bibitem{Leistikow11}
M.D. Leistikow \emph{et al.},
Phys. Rev. Lett. \textbf{107}, 193903 (2011)

\bibitem{Forster1948}
T. F{\"o}rster,
Ann. Phys. \textbf{437}, 55 (1948).

\bibitem{Scholes2003}
G.D. Scholes,
Annu. Rev. Phys. Chem. \textbf{54}, 57 (2003).

\bibitem{Grondelle1994}
R. van Grondelle, J.P. Dekker, T. Gillbro and V. Sundstr{\"o}m,
Biochim. Biophys. Acta. Bioenerg. \textbf{1187}, 1 (1994).

\bibitem{Chanyawadee2009}
S. Chanyawadee, R.T. Harley, M. Henini, D.V. Talapin, and P.G. Lagoudakis,
Phys. Rev. Lett. \textbf{102}, 077402 (2009).

\bibitem{Buhbut2010}
S. Buhbut, S. Itzhakov, E. Tauber, M. Shalom, I. Hod, T. Geiger, Y. Garini, D. Oron, and A. Zaban,
ACS Nano \textbf{4}, 1293 (2010).

\bibitem{Vohra2010}
V. Vohra, G. Calzaferri, S. Destri, M. Pasini, W. Porzio, and C. Botta,
ACS Nano \textbf{4}, 1409 (2010).

\bibitem{Baldo2000}
M.A. Baldo, M.E. Thompson and S.R. Forrest,
Nature (London) \textbf{403}, 750 (2000).

\bibitem{Medintz2003}
I.L. Medintz, A.R. Clapp, H. Mattoussi, E.R. Goldman, B. Fisher, and J.M. Mauro,
Nat. Mater. \textbf{2}, 630 (2003).

\bibitem{Stryer1978}
L. Stryer,
Ann. Rev. Biochem. \textbf{47}, 819 (1978).

\bibitem{Schuler2002}
B. Schuler, E.A. Lipman, and W.A. Eaton,
Nature (London) \textbf{419}, 743 (2002).

\bibitem{Garcia2004}
M.F. Garc{\'i}a-Paraj{\'o}, J. Hernando, G.S. Mosteiro, J.P. Hoogenboom, E.M.H.P. van Dijk, N.F. van Hulst, 
ChemPhysChem \textbf{6}, 819 (2005).

\bibitem{Carriba2008}
P. Carriba, G. Navarro, F. Ciruela, S. Ferr{\'e}, V. Casad{\'o}, L. Agnati, A. Cort{\'e}s, J. Mallol, K. Fuxe, E.I. Canela, C. Llu{\'i}s, and R. Franco,
Nat. Methods \textbf{5}, 727 (2008).


\bibitem{John1991}
S. John and J. Wang,
Phys. Rev. B \textbf{43}, 12772 (1991).


\bibitem{Barenco1995}
A. Barenco, D. Deutsch, A. Ekert, and R. Jozsa,
Phys. Rev. Lett. \textbf{74}, 4083 (1995).

\bibitem{Lovett2003}
B.W. Lovett, J.H. Reina, A. Nazir, and G.A.D. Briggs,
Phys. Rev. B  \textbf{68}, 205319 (2003).

\bibitem{Reina2004}
J.H. Reina, R.G. Beausoleil, T.P. Spiller, and W.J. Munro,
Phys. Rev. Lett. \textbf{93}, 250501 (2004).

\bibitem{Unold2005}
T. Unold, K. Mueller, C. Lienau, T. Elsaesser, and A.D. Wieck,
Phys. Rev. Lett. \textbf{94}, 137404 (2005).

\bibitem{Lakowicz2006}
J.R. Lakowicz, 
\emph{Principles of Fluorescence Spectroscopy} (Springer, Berlin, 2006).

\bibitem{Sprik1996}
R. Sprik, B.A. van Tiggelen, and A. Lagendijk,
Europhys. Lett. \textbf{35}, 265 (1996).

\bibitem{Barnes1998}
W.L. Barnes,
J. Mod. Opt. \textbf{45}, 661 (1998).





\bibitem{Yang2008}
Z.W. Yang, X.R. Zhou, X.G. Huang, J. Zhou, G. Yang, Q. Xie, L. Sun, and B. Li,
Opt. Lett. \textbf{33}, 1963 (2008).

\bibitem{Kolaric2007}
B. Kolaric, K. Baert, M. van der Auweraer, R.A.L. Vall{\'e}e, and K. Clays,
Chem. Mater. \textbf{19}, 5547 (2007).

\bibitem{Andrew2000}
P. Andrew and W.L. Barnes,
Science \textbf{290}, 785 (2000).

\bibitem{Nakamura2005}
T. Nakamura, M. Fujii, K. Imakita, and S. Hayashi,
Phys. Rev. B \textbf{72}, 235412 (2005).

\bibitem{Dung2002}
H.T. Dung, L. Kn{\"o}ll, and D.-G. Welsch, 
Phys. Rev. A \textbf{65}, 043813 (2002).

\bibitem{Nakamura2006}
T. Nakamura, M. Fujii, S. Miura, M. Inui, and S. Hayashi,
Phys. Rev. B \textbf{74}, 045302 (2006).

\bibitem{Dood2005}
M.J.A. de Dood, J. Knoester, A. Tip, and A. Polman,
Phys. Rev. B \textbf{71}, 115102 (2005).

\bibitem{Chance1978}
R.R. Chance, A. Prock, and R. Silbey,
Adv. Chem. Phys. \textbf{37}, 1 (1978).

\bibitem{Rindermann2011}
J.J. Rindermann, Y. Akhtman, J. Richardson, T. Brown, and P.G. Lagoudakis,
J. Am. Chem. Soc. \textbf{133}, 279 (2011).

\bibitem{footnote:EPAPS}
See EPAPS document $``EPAPS\_Blum\_supplement.pdf``$ for theoretical derivation, experimental details, data analysis, see www.aip.org/pubservs/epaps.html.

\bibitem{footnote:reference}
In the reference sample the donor-only fluorescence decay rate is found to be $\gamma_{\rm D}^0$$=$$0.367$ $\pm 0.004$  ns$^{-1}$. 
Adding the FRET acceptor results in the increase of the donor decay to $\gamma_{\rm DA}^0 = 0.555 \pm 0.009$ ns$^{-1}$. 
Using Eq.~(\ref{gammaFRET}) we find a reference FRET rate $\gamma_{\rm FRET}^{0}= 0.188 \pm 0.010$ ns$^{-1}$.
Using Eq.~(\ref{etaFRET}) we find a reference energy transfer efficiency $\eta_{\rm FRET}^{0} = 34 \pm 3 \%$, in reasonable agreement with the value of $28\%$ estimated from emitter spectral properties, distances and orientations~\cite{Rindermann2011, footnote:EPAPS, Norman2000}. 
From the efficiency we also conclude that the distance between donor and acceptor as set by the DNA-strand is close to the F{\"o}rster distance $R_{0}$. 

\bibitem{Cesa2009}
Y. Cesa, C. Blum, J.M. van den Broek, A.P. Mosk, W.L. Vos, and V. Subramaniam,
Phys. Chem. Chem. Phys. \textbf{11}, 2525 (2009).

\bibitem{Norman2000}
D.G. Norman, R.J. Grainger, D. Uhrin, and D.M.J. Lilley,
Biochem. \textbf{39}, 6317 (2000).



\end{thebibliography}
\end{document}